\documentclass[preprint,notoc]{JHEP3}
\usepackage{epsfig}

\def\eslt{E_T^{\rm miss}}

\def\esl{\not\!\!{E}}
\def\psl{\not\!\!{p}}
\def\msl{\not\!\!{m}}

\def\to{\rightarrow}

\def\bi{\begin{itemize}}
 \def\ei{\end{itemize}}
\def\te{\tilde e}

\def\c1p{C1^\prime}
\def\ta{\tilde a}
\def\tG{\tilde G}

\def\tu{\tilde u}

\def\ta{\tilde a}

\def\tb{\tilde b}

\def\tst{\tilde t}

\def\tg{\tilde g}

\def\tq{\tilde q}

\def\tw{\widetilde W}
\def\tz{\widetilde Z}
\def\alt{\stackrel{<}{\sim}}
\def\agt{\stackrel{>}{\sim}}
\def\be{\begin{equation}}  
\def\ee{\end{equation}}  
\def\bea{\begin{eqnarray}}  
\def\eea{\end{eqnarray}}  

\def\sps1ap{SPS1a$^\prime$}
\newcommand\annp[3]{{\it Annals\ Phys.\ }{\bf #1} (#2) #3}
\newcommand\sjp[3]{{\it Sov.\ J.\ Nucl.\ }{\bf #1} (#2) #3}
\newcommand\njp[3]{{\it New\ J.\ Phys.\ }{\bf #1} (#2) #3}

\title{Hidden SUSY at the LHC:\\
the light higgsino-world scenario\\
and the role of a lepton collider
}
\author{Howard Baer$^{a}$, Vernon Barger$^b$ and Peisi Huang$^b$ \\
$^a$Dept.\ of Physics and Astronomy, University of Oklahoma, Norman, OK 73019, USA\\
$^b$Dept. of Physics, University of Wisconsin, Madison, WI 53706, USA\\
E-mail: \email{baer@nhn.ou.edu}, \email{barger@pheno.wisc.edu},
\email{phuang7@wisc.edu}}

\preprint{\vbox{}}

\abstract{
While the SUSY flavor, CP and gravitino problems seem to favor a very heavy
spectrum of matter scalars, fine-tuning in the electroweak sector
prefers low values of superpotential mass $\mu$.
In the limit of low $\mu$, the two lightest neutralinos and light chargino
are higgsino-like.
The light charginos and neutralinos may have large production cross sections at LHC,
but since they are nearly mass degenerate, there is 
only small energy release in three-body sparticle decays.
Possible dilepton and trilepton signatures are difficult to 
observe after mild cuts due to the very soft $p_T$ spectrum
of the final state isolated leptons. 
Thus, the higgsino-world scenario can easily elude standard SUSY searches
at the LHC. It should motivate experimental searches to focus on
dimuon and trimuon production at the very lowest $p_T(\mu )$ values possible.
If the neutralino relic abundance is enhanced via non-standard cosmological dark matter production, 
then there exist excellent prospects for direct or indirect detection of
higgsino-like WIMPs.
While the higgsino-world scenario may easily hide from LHC SUSY searches, 
a linear $e^+e^-$ collider or a muon collider operating in the $\sqrt{s}\sim 0.5-1$ TeV
range would be able to easily access the chargino and 
neutralino pair production reactions.
}  
\keywords{Supersymmetry
Phenomenology, Supersymmetric Standard Model, Large Hadron Collider}

\begin{document}

\section{Introduction}
\label{sec:intro}

The well-known instability of the scalar sector of the Standard Model (SM) 
to quadratic divergences is elegantly solved by the introduction
of supersymmetry (SUSY)\cite{review}. 
In the case of the Minimal Supersymmetric
Standard Model (MSSM) with soft SUSY breaking (SSB) terms, the divergences
in the scalar sector are rendered to merely logarithmic.
Interplay between the electroweak sector and the SUSY partners suggests 
the superpartner masses should exist at or around the TeV scale to avoid
re-introduction of fine-tuning.

While the MSSM may be very appealing, it does suffer several pathologies.
Unfettered soft SUSY breaking terms lead to large rates for flavor-changing
neutral current processes and $CP$ violation\cite{masiero}.
Inclusion of grand unified theories with SUSY may lead to unacceptably 
high rates for proton decay\cite{pdecay}. 
And in gravity-mediated SUSY breaking models (SUGRA), 
gravitino production followed by late-time gravitino decays
in the early universe are in conflict with the successful picture of Big Bang 
nucleosynthesis (BBN) unless re-heat temperatures after inflation are limited
to $T_R\alt 10^5$ GeV\cite{gravprob}. 
The latter bound is in conflict with appealing
baryogenesis models such as thermal\cite{leptog} (or non-thermal\cite{ntlepto}) leptogenesis, 
which require $T_R\agt 2\times 10^9$ GeV ($10^6$ GeV).

A common solution to the above four problems is to push the SUSY matter
scalars into the multi-TeV regime\cite{dine}. 
The heavy scalars thus suppress loop-induced
flavor and $CP$ violating processes, and suppress proton decay rates.
If the multi-TeV scalars derive from a SUGRA model with a simple form
for the K\"ahler potential, 
then the gravitino mass $m_{3/2}$ is also expected to
exist in the multi-TeV range. By pushing $m_{3/2}$ into the 10-50 TeV
range, the gravitino lifetime can be reduced to $\tau_{3/2}\alt 1$ second,
so the gravitino decays shortly before BBN begins, this solving the 
gravitino problem\cite{moroi}.

At first glance, multi-TeV gravitino and scalar masses seem in conflict with
SUSY electroweak fine-tuning. The possible SUSY electroweak fine-tuning 
arises from minimization of the scalar potential after electroweak 
symmetry breaking. Here, the tree-level electroweak breaking conditions
are familiarly written as\cite{wss}
\bea
B\mu=\frac{(m_{H_u}^2+m_{H_d}^2+2\mu^2 )\sin 2\beta}{2}\;, 
\label{eq:mssmB}\ \ \ {\rm and}\\
\mu^2 =\frac{m_{H_d}^2-m_{H_u}^2\tan^2\beta}{(\tan^2\beta -1)}
-\frac{M_Z^2}{2}  ,
\label{eq:mssmmu}
\eea
where $B$ is the bilinear SSB term and $m_{H_u}^2$ and $m_{H_d}^2$
are the up and down Higgs SSB masses evaluated at the weak scale,
$\mu$ is the superpotential Higgs mass term and $\tan\beta$ is the ratio
of Higgs field vevs: $\tan\beta ={v_u\over v_d}$. 

A measure of fine-tuning 
\be
\Delta_i\equiv\left|\frac{\partial\log M_Z^2}{\partial\log a_i}\right|
\ee 
was advocated in Ref. \cite{barbieri}.
More sophisticated measures were advocated in Ref's \cite{diego},
while in Ref. \cite{ccn}, the $\mu$ parameter itself is taken as a measure
of fine-tuning: the latter paper requires $|\mu |\alt 1$ TeV to avoid 
too much fine-tuning. This measure motivates the well-known hyperbolic branch/
focus point (HB/FP) region of minimal supergravity (mSUGRA or CMSSM) as
allowing for heavy scalars with low $\mu$ value and low 
fine-tuning\cite{ccn,feng}. A virtue of the HB/FP region is that 
multi-TeV scalars can co-exist with apparent low levels of
electroweak fine-tuning.

In this paper, we will consider supersymmetric models with
large, multi-TeV scalar masses, but with low, sub-TeV 
superpotential $\mu$ term. We consider the case with
intermediate range gaugino masses. This scenario, with
\be
|\mu |\ll m_{gauginos}\ll m_{scalars} ,
\ee
has been dubbed ``higgsino-world'' by Kane\cite{kane}, and leads to
a sparticle mass spectrum with a light higgsino-like chargino $\tw_1$ and
two light higgsino-like neutralinos $\tz_1$ and $\tz_2$. In models
with gaugino mass unification at $M_{GUT}$, then the state
$\tz_3$ will be mainly bino-like, while $\tz_4$ and $\tw_2$ 
will be wino-like. 

While the higgsino-world scenario seems highly appealing due to its ability to reconcile
multi-TeV scalars and gravitinos with low electroweak fine-tuning, it has perhaps fallen
out of favor for two reasons. First, higgsino-world leads to a very low thermal relic density of
neutralinos, not at all in accord with measurements from WMAP and
other experiments which require\cite{wmap7}
\be
\Omega_{\rm DM}h^2=0.1123\pm 0.0035\ \ \ {\rm at\ 68\%\ CL} .
\ee
Second, higgsino-world scenarios are not easily realized in the paradigm mSUGRA/CMSSM framework, 
since elevating scalar masses into the multi-TeV region for a given value of GUT scale 
gaugino mass $m_{1/2}$ pushes one beyond the HB/FP region into a portion of
parameter space where radiative EWSB is not 
realized under the assumption of universal scalar masses $m_0$ at $M_{GUT}$.

Pertaining to the dark matter issue, 
a number of recent works have emphasized that the standard
picture of a thermal SUSY WIMP as dark matter is subject to
very high fine-tuning\cite{bbox}. 
Furthermore, non-standard cosmologies have many desirable features, 
and may even be favored by string theoretic constructions. 
For instance, Kane {\it et al.} have shown\cite{lightmod} that
at least one moduli field in string theory should maintain
a mass at or around the 10 TeV scale. Such moduli fields can be produced
via coherent oscillations in the early universe, and decay into WIMPs,
thereby augmenting the WIMP abundance\cite{mr}, 
or they can decay into SM particles, thus 
generating entropy and diluting the WIMP abundance. 
Gelmini {\it et al.} have shown in this case that
SUSY models with {\it any} value-- either too high or too low-- of thermal WIMP abundance 
may give rise to the measured CDM abundance via the enhancement or diminution 
due to scalar field (moduli) decays\cite{gg}.
In particular, for higgsino-world with too low a thermal WIMP abundance, 
the light higgsino abundance can be enhanced
by moduli decays, leading to the correct abundance of higgsino-like WIMPs.

Alternatively, in SUSY models wherein the strong $CP$ problem is solved by the
introduction of Peccei-Quinn-Weinberg-Wilczek\cite{pqww} 
invisible axion\cite{ksvz,dfsz}, one must introduce an axion
supermultiplet, which contains an $R$-parity even spin-0 saxion $s(x)$, 
along with an $R$-parity odd spin-${1\over 2}$ axino $\ta (x)$, in
addition to the light pseudoscalar axion field $a(x)$. In models such as these,
with a TeV-scale axino and a higgsino-like neutralino as LSP, the
$\tz_1$ abundance can be augmented by axino production and subsequent
re-annihilation at temperatures above BBN but below 
neutralino freeze-out\cite{ckls}.
Depending on the various PQMSSM model parameters, the CDM consists
of an axion/neutralino admixture, where either the axion or the
neutralino can dominate the abundance\cite{blrs}.

A third modifiation of the thermal WIMP abundance may also occur: 
a model with a supposed underabundance of neutralino dark matter 
may enjoy enhancement of the relic DM
abundance due to thermal gravitino production\cite{gravprod} followed by 
cascade decays to the LSP state\cite{grav,bdrs}. 

Pertaining to the issue of higgsino-world being difficult to realize in 
the paradigm mSUGRA model, we note that it is easily realized in models
with non-universal GUT scale Higgs masses (NUHM)\cite{nuhm}. In fact, in GUT models
such as $SO(10)$, the matter supermultiplets live in the 16-dimensional
spinor representation, while Higgs superfields live in 10 or other dimensional multiplets.
In such models, there is little reason to expect matter-Higgs SSB universality at $M_{GUT}$.

For the above reasons, we feel that it may be
opportune to reconsider the higgsino-world scenario, and whether such a scenario
would be visible to LHC SUSY searches.
Toward this end, we discuss in 
Sec. \ref{sec:pspace} the higgsino-world parameter space and expected mass spectra.
In Sec. \ref{sec:Oh2}, we present calculations of the standard thermal neutralino abundance in
the higgsino-world scenario, and discuss its direct and indirect detection in the case where
non-standard cosmological processes augment the relic higgsino abundance.
In Sec. \ref{sec:lhc}, we evaluate the dominant sparticle production cross sections for 
the lighter matter states at the LHC and calculate their branching fractions. 
While higgsino production cross sections occur at possibly observable levels, 
the compressed spectra lead to sparticle decays with very low energy release, 
and very soft detectable particles. 
To the best of our knowledge, higgsino-world SUSY can effectively elude standard SUSY searches
for jets plus missing $E_T$ $(MET)$, 
and also for isolated multi-leptons$+MET$ at LHC7 (LHC at $\sqrt{s}=7$ TeV) 
with $\sim 10$ fb$^{-1}$ of integrated luminosity.
In Sec. \ref{sec:ilc}, we discuss higgsino-world signatures at a TeV scale 
lepton collider such as ILC or a muon collider (MC).
In Sec. \ref{sec:conclude}, we present our final discussion and 
conclusions.

\section{Higgsino-world parameter space and mass spectra}
\label{sec:pspace}

We will adopt the Isajet 7.81 program for SUSY particle mass spectrum generation\cite{isajet}.
To generate spectra in higgsino-world scenario, we will adopt the 
Isasugra non-universal Higgs mass parameter space (NUHM2):
\be
m_0,\ m_{1/2},\ A_0,\ \tan\beta ,\ \mu ,\ m_A .
\ee
In the above parameter space, $m_0$, $m_{1/2}$ and $A_0$ are the usual
GUT scale parameters, although here $m_0$ is reserved only for matter 
scalars, and not Higgs scalar soft masses. 
The two additional parameters $\mu$ and $m_A$
are stipulated at the weak scale, and are used to solve for the weak
scale values of $m_{H_u}^2$ and $m_{H_d}^2$. These latter parameters
are run from the weak to GUT scale, and their GUT scale values are
determined by enforcing the input weak scale values of $\mu$ and $m_A$.
We will take $m_0\sim m_{\tG}$ to be in the multi-TeV range, so that we obtain a 
decoupling solution to the SUSY flavor, CP, $p$-decay and gravitino problems.
Thus, the parameters $m_0$ and also $A_0$ and $m_A$ will be largely irrelevant 
for our analysis. The main parameter space dependence will arise from just varying 
$\mu$ and $m_{1/2}$. Since we are interested in the light higgsino-world scenario, 
with $\mu \ll M_i$ (where $M_i$ are the weak scale gaugino masses), the parameter 
$\tan\beta$, which induces gaugino-higgsino mixing, will also not be terribly relevant.

The higgsino-world input parameters and mass spectra for two sample benchmark
points with $\mu =150$ and 300 GeV are listed in Table \ref{tab:BM}. 
 We also take $m_0=5000$ GeV, $m_{1/2}=800$ GeV, $A_0=0$, $\tan\beta =10$
and $m_A=800$ GeV. 
The spectra are also shown in Fig. \ref{fig:mass} for the higgsino-world 
case where $\mu =150$ GeV (HW150).
%
\begin{table}\centering
\begin{tabular}{lcc}
\hline
parameter & HW150 & HW300 \\
\hline
$m_0$      & 5000 & 5000 \\
$m_{1/2}$  & 800 & 800 \\
$A_0$      & 0 & 0 \\
$\tan\beta$& 10 & 10 \\
$\mu$      & 150 & 300  \\
$m_A$      & 800 & 800 \\
\hline
$m_{\tg}$   & 2004.9 & 2004.2   \\
$m_{\tu_L}$ & 5171.5 & 5171.4   \\
$m_{\tst_1}$& 3240.2 & 3243.8   \\
$m_{\tb_1}$ & 4267.8 & 4269.4  \\
$m_{\te_R}$ & 4869.4 & 4870.1  \\
$m_{\tw_2}$ & 672.7  & 675.4  \\
$m_{\tw_1}$ & 156.3  & 310.5  \\
$m_{\tz_4}$ & 688.2 & 691.0  \\ 
$m_{\tz_3}$ & 356.3 & 366.9  \\ 
$m_{\tz_2}$ & 158.9 & 311.4  \\ 
$m_{\tz_1}$ & 142.7 & 283.2  \\ 
$m_h$       & 120.1 & 120.1   \\ 
\hline
$\sigma (LHC7)$ & 1055 fb & 63.5 fb \\
$\Omega_{\tz_1}^{std}h^2$ & 0.008 & 0.03 \\
$BF(b\to s\gamma)$ & $3.5\times 10^{-4}$  & $3.5\times 10^{-4}$  \\
$\sigma^{SI}(\tz_1 p)$ (pb) & $1.0\times 10^{-8}$  & $3.1\times 10^{-8}$ \\
$\langle\sigma v\rangle |_{v\to 0}$  (cm$^3$/sec) 
& $0.28\times 10^{-24}$  & $0.09\times 10^{-24}$ \\
$v_H^{(1)}$ & 0.98 & 0.90 \\
\hline
\end{tabular}
\caption{Input parameters and masses in~GeV units
for two higgsino-world scenario benchmark points 
HW150 and HW300, with $\mu =150$ and 300 GeV, respectively.
}
\label{tab:BM}
\end{table}

From Table \ref{tab:BM} or Fig. \ref{fig:mass}, we see that the three states $\tw_1$, 
$\tz_1$ and $\tz_2$ all have masses clustered around the value
of $\mu =150$ or 300 GeV. These states are dominantly higgsino-like.
The weak scale gaugino masses $M_1\sim 352$ GeV and $M_2\sim 638$ GeV
for the two cases, so that $\tz_3$ is bino-like and
$\tw_2$ and $\tz_4$ are wino-like. The squarks and sleptons all 
are decoupled, with masses in the multi-TeV range, since $m_0=5$ TeV.
The $\tz_2-\tz_1$ mass gap is just 16.2 GeV and 28.2 GeV, 
respectively, for the two cases. 
We also show the higgsino fraction of the lightest neutralino:
$v_H=\sqrt{v_1^{(1)2}+v_2^{(1)2}}$ where $v_1^{(i)}$ is the
higgsino $\tilde{h}_u^0$ content and $v_2^{(i)}$ is the
higgsino $\tilde{h}_d^0$ content of neutralino $\tz_i$ in the
notation of Ref. \cite{wss}. Here, the value of $v_H$ is
$0.98$ for HW150 and $0.9$ for HW300. Increasing
$\mu$ to 500 GeV, using the same choice of other model parameters,
decreases $v_H\sim 0.21$, 
so that in this case the lightest neutralino is no longer
dominantly higgsino-like, but rather of mixed higgsino-bino variety.
We also see from Table \ref{tab:BM} that the standard
thermal neutralino abundance is $\Omega_{\tz_1}^{std}h^2\sim 0.008$
and $0.03$, repectively, {\it i.e.} well below the WMAP-measured CDM abundance.
\FIGURE[tbh]{
\includegraphics[width=13cm,clip]{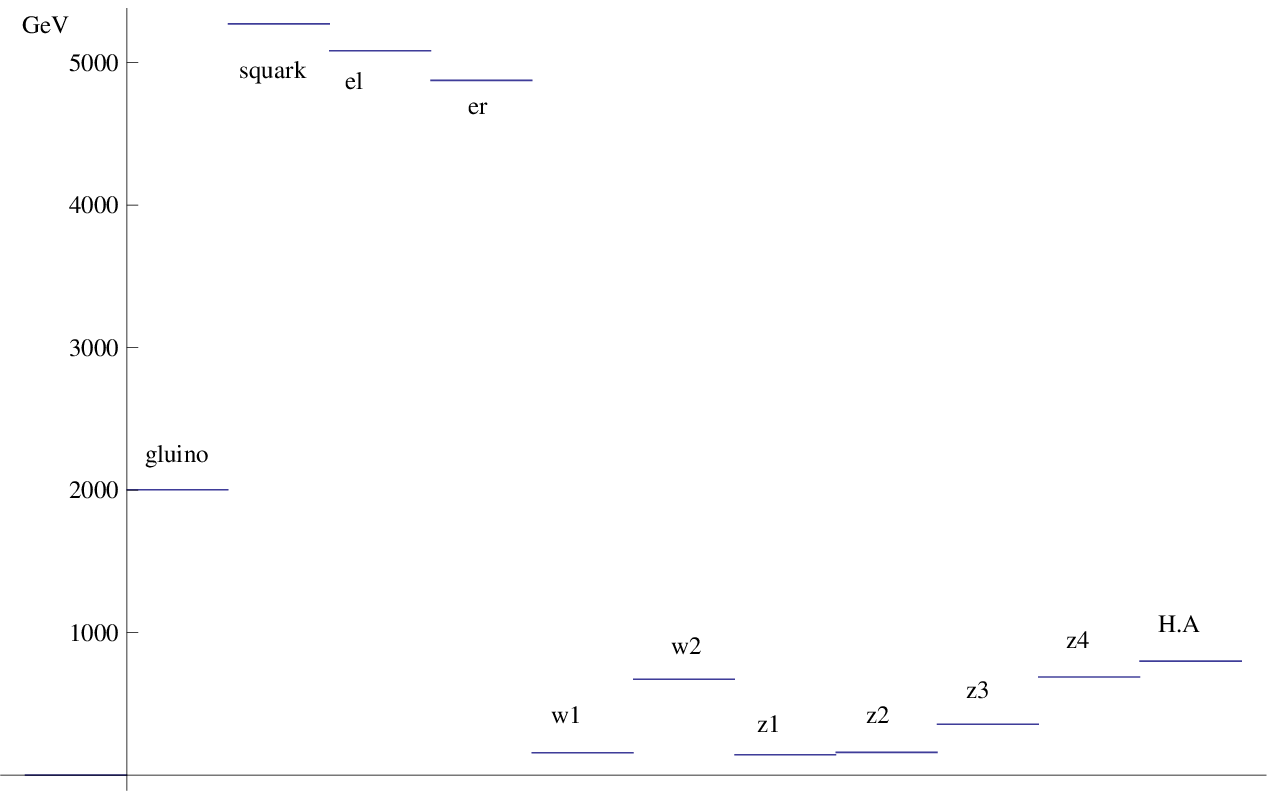}
\caption{Sparticle mass spectra for a light higgsino world 
scenario with $\mu =150$ GeV, $m_A=800$ GeV, 
$m_0=5000$ GeV, $m_{1/2}=600$ GeV, $A_0=0$ and $\tan\beta =10$.
We take $m_t=173.3$ GeV.
}
\label{fig:mass}}

In Fig. \ref{fig:vH}, we show color-coded contours of the higgsino fraction $v_H$ 
of the lightest neutralino $\tz_1$ in the $\mu\ vs.\ m_{1/2}$ parameter space
plane for $m_0=5$ TeV, $A_0=0$ and $\tan\beta =10$.
The green, yellow and especially red regions contain a lightest neutralino with
large higgsino fraction $v_H\agt 0.5$. This region essentially defines the
higgsino-world scenario parameter space, which is found at low $|\mu |$
and large $m_{1/2}$. As one enters the blue-shaded region, the
$\tz_1$ becomes increasingly bino-like. The unshaded region at low $\mu$ 
is excluded by LEP2 limits on the lightest chargino: $m_{\tw_1}\alt 103.5$ GeV.
\FIGURE[tbh]{
\includegraphics[width=13cm,clip]{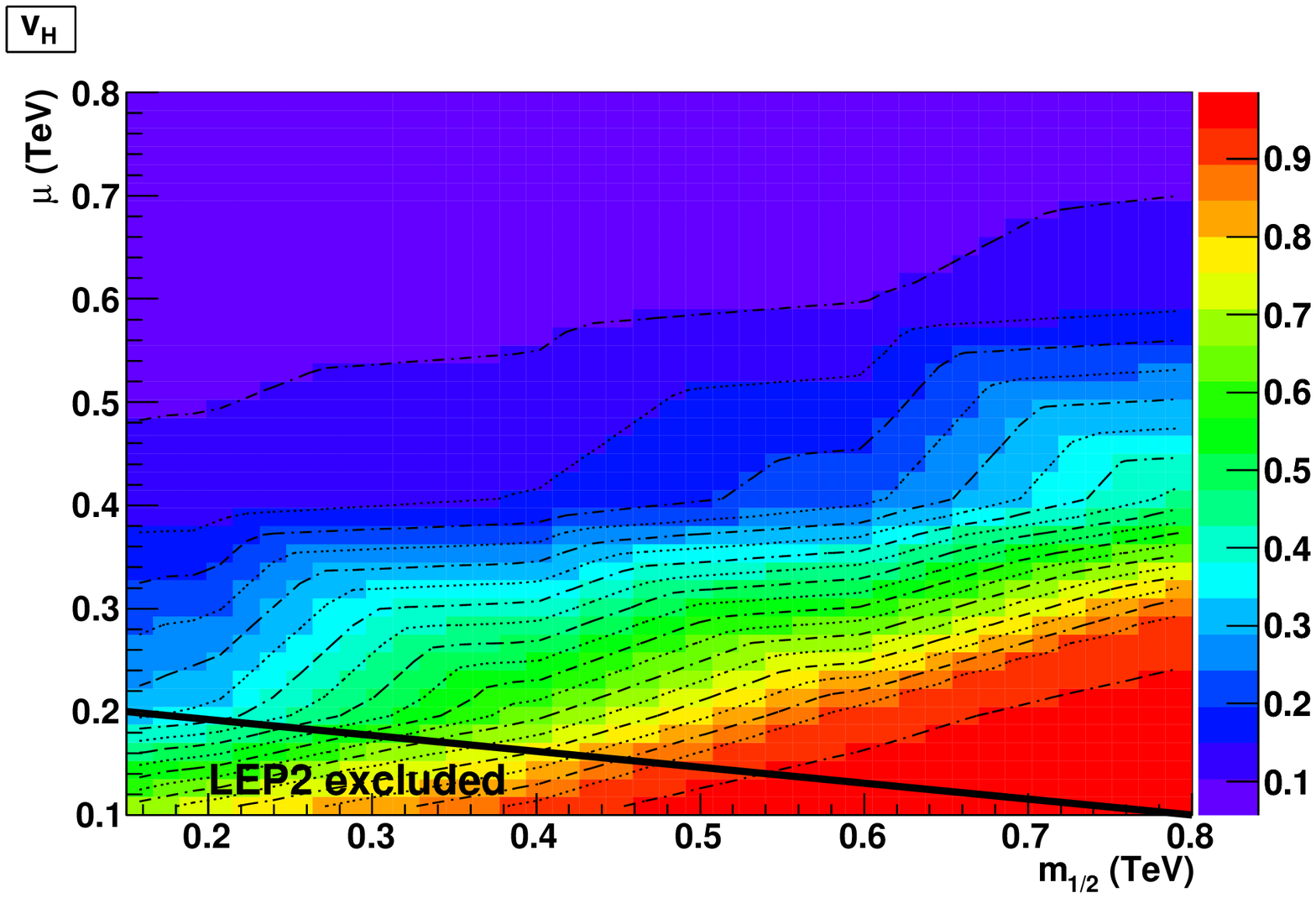}
\caption{Color-coded contours of higgsino content $v_H$ of the $\tz_1$
in the $\mu\ vs.\ m_{1/2}$ plane with $m_A=800$ GeV, 
$m_0=5000$ GeV, $A_0=0$ and $\tan\beta =10$.
}
\label{fig:vH}}

In Fig. \ref{fig:mz1}, we show {\it a}). the color-coded mass contours of the lightest neutralino $\tz_1$, 
and {\it b}). the $m_{\tz_2}-m_{\tz_1}$ mass gap (which is always very close to the value of the
$m_{\tw_1}-m_{\tz_1}$ mass gap). In the higgsino-world scenario
with low $\mu$ and large $m_{1/2}$, we find that $m_{\tz_1}$ can drop as low
as $\sim 90$ GeV, where the lower limit comes from the LEP2 constraint on chargino masses. 
Meanwhile, the mass gap $m_{\tz_2}-m_{\tz_1}$ drops as low as $\sim 10$
GeV in the extreme higgsino-world region. Thus, for extreme higgsino-world
parameters, we always expect the $\tz_2$ states to decay via three-body modes or
loop-suppressed two-body decays such as $\tz_2\to \tz_1\gamma$\cite{z2z1g}. Two-body decays
such as $\tz_2\to\tz_1 Z$ or $\tz_1 h$ will always be closed in the 
higgsino-world scenario.
\FIGURE[tbh]{
\includegraphics[width=13cm,clip]{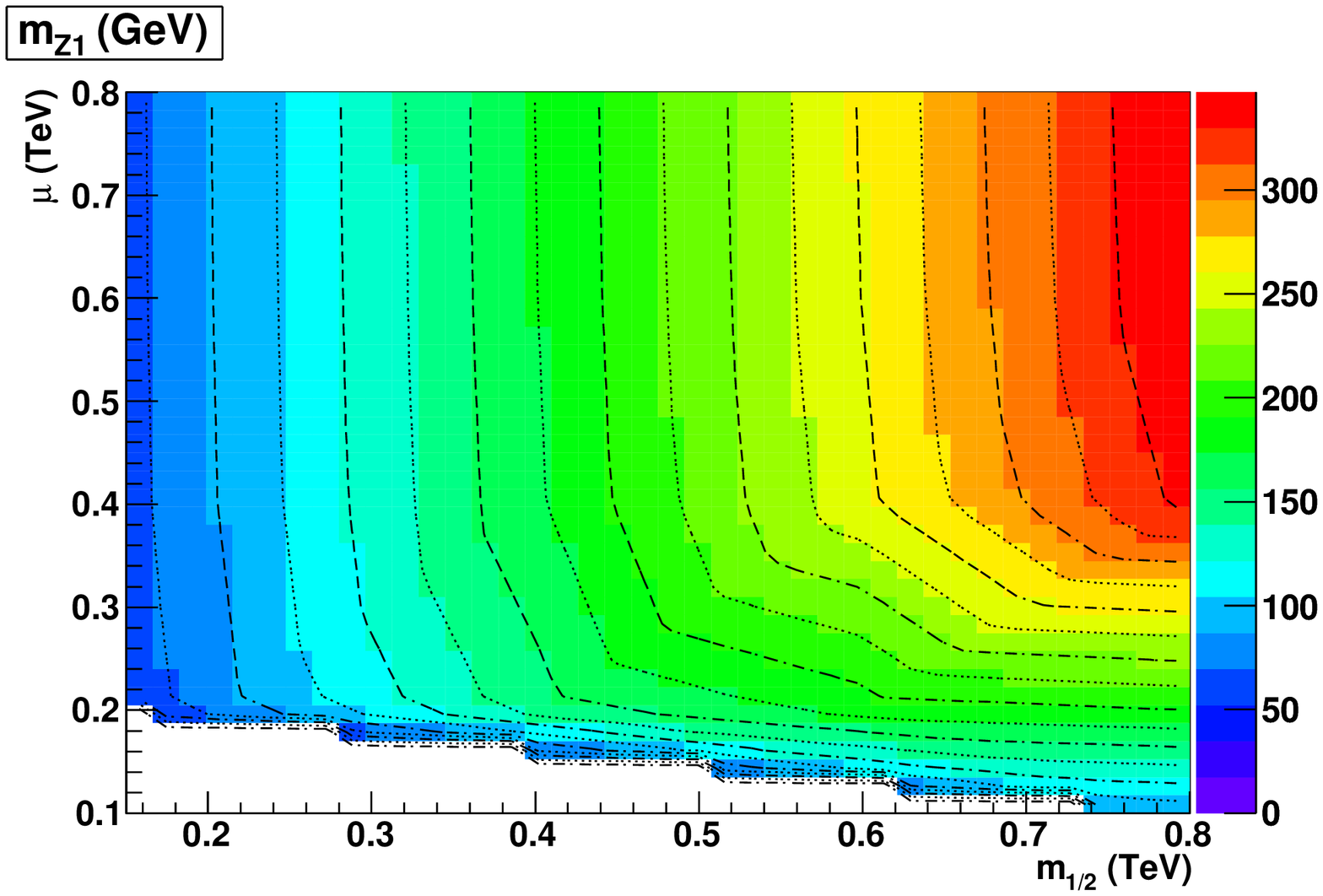}
\includegraphics[width=13cm,clip]{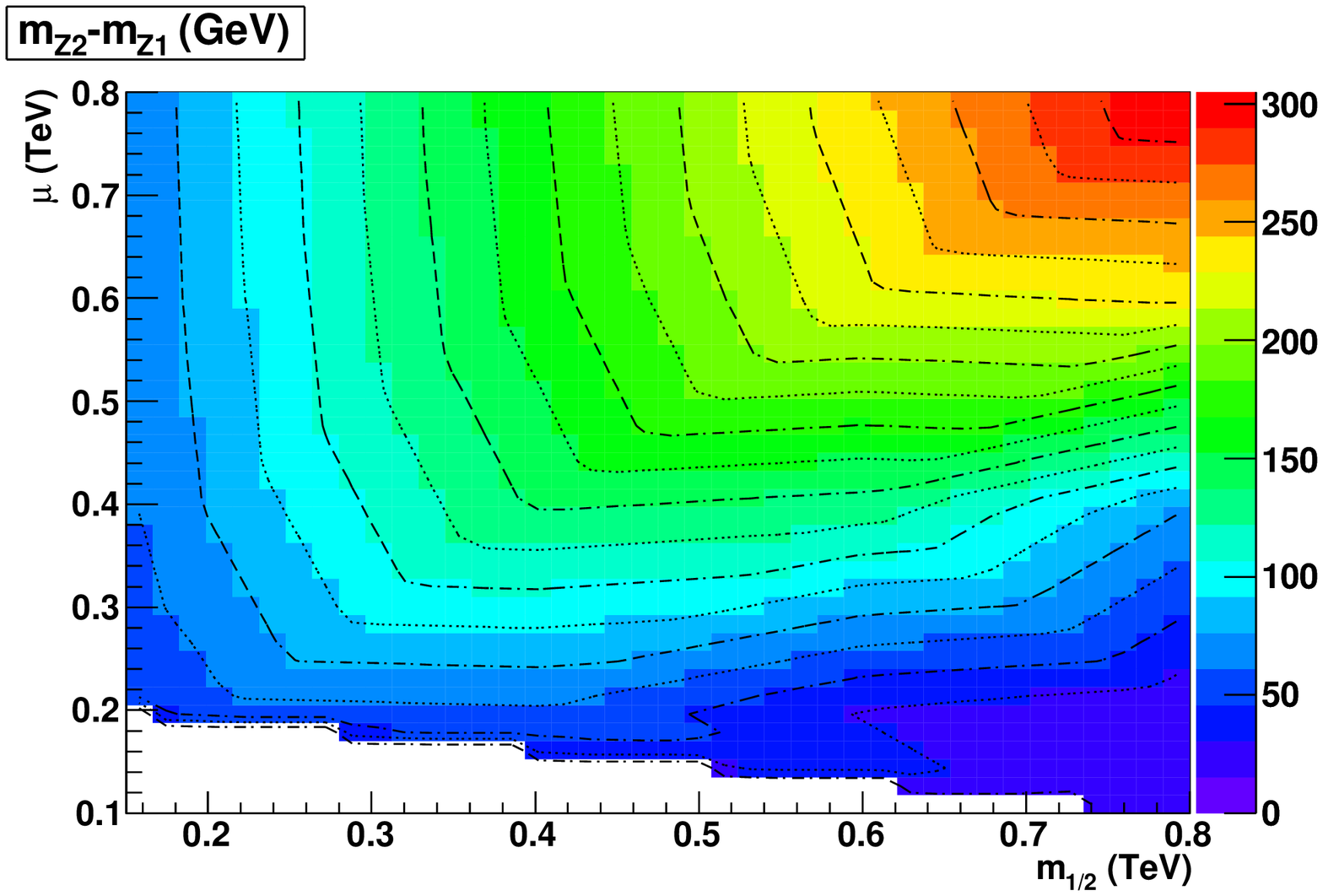}
\caption{In {\it a})., we plot contours of $m_{\tz_1}$ while in {\it b}). we plot
contours of $m_{\tz_2}-m_{\tz_1}$ 
in the $\mu\ vs.\ m_{1/2}$ plane with $m_A=800$ GeV, 
$m_0=5000$ GeV, $A_0=0$ and $\tan\beta =10$.
}
\label{fig:mz1}}
%

\section{Neutralino relic density and direct detection rates}
\label{sec:Oh2}

In the light higgsino-world scenario, if the lightest
neutralino is dominantly higgsino-like, then it will
sustain large $\tz_1\tz_1$ annihilation cross sections 
into vector boson states $ZZ$ and $W^+W^-$. This will
result in a standard thermal neutralino
abundance typically well below WMAP-measured values
of $\Omega_{CDM}h^2\sim 0.11$. In Fig. \ref{fig:oh2}, we
show color-coded contours of the standard thermal
neutralino abundance $\log_{10}\Omega_{\tz_1}^{std}h^2$.
We see that indeed in the low $\mu$ region $\Omega_{\tz_1}^{std}h^2$
is at the $10^{-2.5}-10^{-1}$ range, where $10^{-1}$
occurs for mixed higgsino-bino states. Thus, a non-standard
cosmology is likely needed to explain the CDM abundance in
the higgsino-world scenario. Indeed, many ``non-standard''
scenarios can be highly motivated by other physics considerations
(the presence of TeV-scale moduli in string theory, the axion solution to the
strong CP problem $\cdots$), and so may well be more appealing
than simple thermal production of WIMPs.
\FIGURE[tbh]{
\includegraphics[width=13cm,clip]{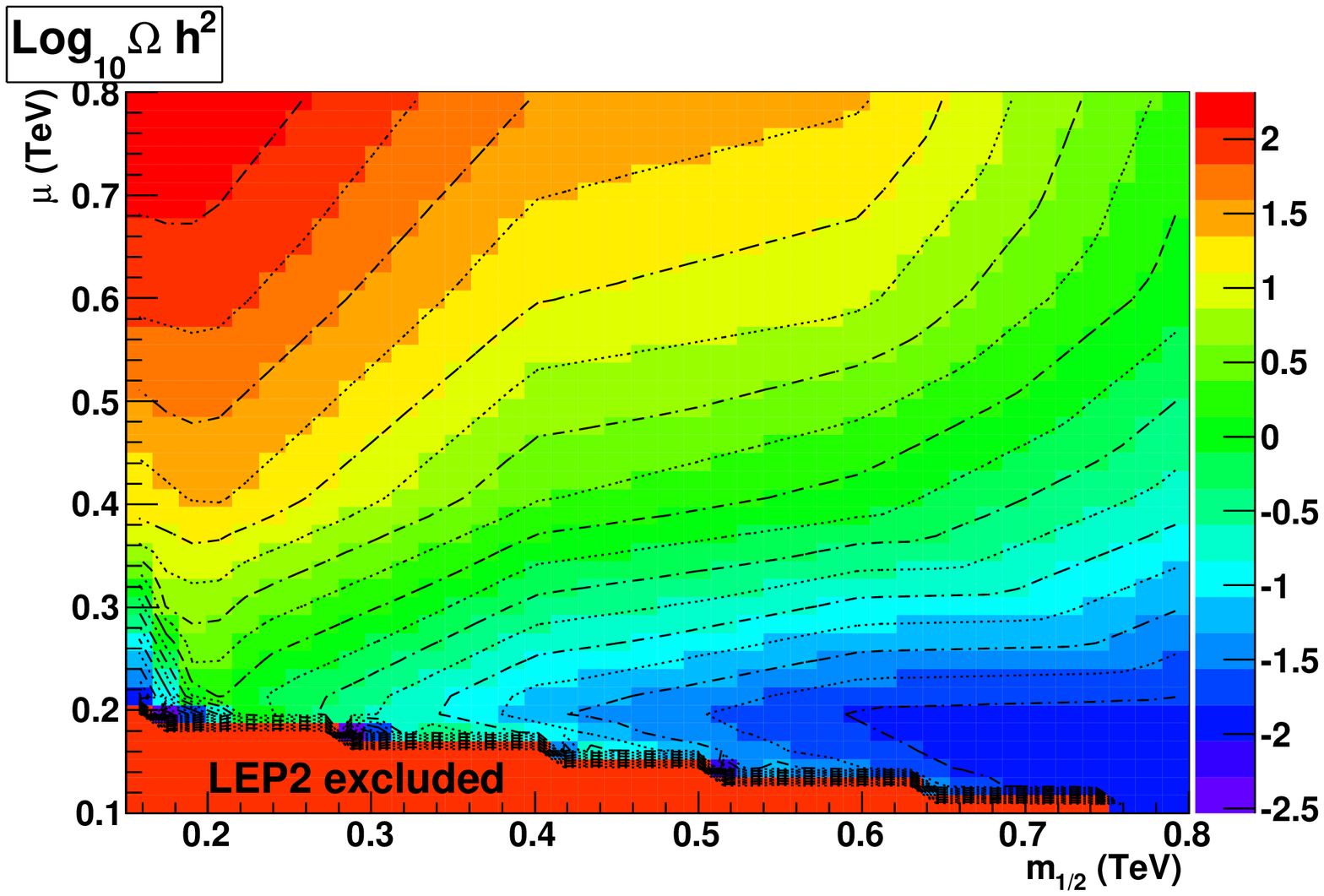}
\caption{Regions of thermal neutralino abundance 
in the $\mu\ vs.\ m_{1/2}$ plane with $m_A=800$ GeV, 
$m_0=5000$ GeV, $A_0=0$ and $\tan\beta =10$.
}
\label{fig:oh2}}

If the higgsino relic abundance is enhanced (say, by moduli decays, or by axino
production and decay, or by gravitino production and decay) beyond standard
expectations, then a higgsino-like WIMP may well make up the bulk of dark matter.
In this case, we present in Fig. \ref{fig:sigSI} color-coded contours of
spin-independent neutralino-proton scattering cross section in
units of $10^{-9}$ pb in the $\mu\ vs.\ m_{1/2}$ plane. The red and yellow shaded
regions have $\sigma_{SI}(p\tz_1)\agt 30\times 10^{-9}$ pb, while green-shaded
regions have $\sigma_{SI}(p\tz_1)\agt 20\times 10^{-9}$ pb. The Xenon-100
experiment\cite{xe100}-- for $m_{WIMP}\sim 100-200$ GeV--excludes 
$\sigma_{SI}(\tz_1 p)\agt 10^{-8}$ pb, so already a large portion
of higgsino-world parameter space is excluded if higgsino-like
WIMPs make up all the CDM. It is shown in Ref. \cite{blrs} that in
the case of the Peccei-Quinn augmented MSSM, where an axion-axino-saxion
supermultiplet is required to solve the strong $CP$-problem, that for
some ranges of PQMSSM parameters, higgsino-like WIMPs could make up 
virtually all the DM abundance, while for other PQMSSM parameters, 
(low re-heat temeprature $T_R$ or if cosmologically produced axinos
decay to neutralinos before neutralino freeze-out), then the CDM abundance
may be axion-dominated, while the higgsino abundance maintains its
standard relic density. In this latter case, the assumed WIMP abundance would have
to be scaled down by a factor of 10-100, and so the higgsino-world scenario
would then escape Xenon-100 null-search constraints.
\FIGURE[tbh]{
\includegraphics[width=13cm,clip]{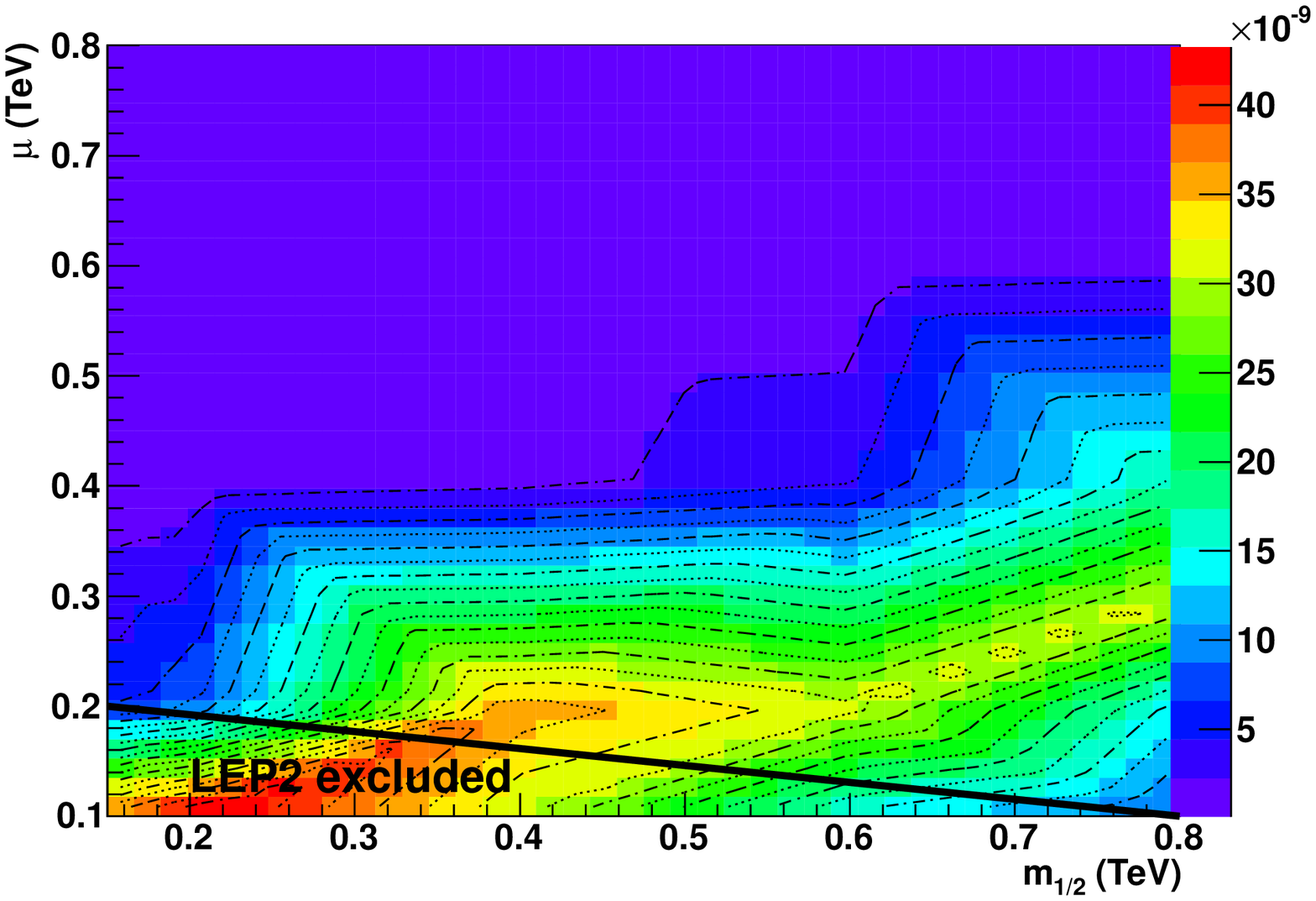}
\caption{Contours of $\sigma^{SI}(\tz_1 p)$
in units of $10^{-9}$ pb in the $\mu\ vs.\ m_{1/2}$ plane with $m_A=800$ GeV, 
$m_0=5000$ GeV, $A_0=0$ and $\tan\beta =10$.
}
\label{fig:sigSI}}

If higgsino-like WIMPs comprise the bulk of dark matter, then it may also
be possible to detect them via searches for galactic halo WIMP annihilation into
final states containing positrons, anti-protons, gamma-rays or anti-deuterons\cite{bbko}.
In these cases, the WIMP annihilation rate is always proportional
to thermally averages WIMP annihilation cross section times relative
velocity, in the limit where $v\to 0$ (in the galactic halo): 
$\langle\sigma v\rangle|_{v\to 0}$. The exact detection rates will also depend
on various astrophysical quantities, and details of the detection devices and their
backgrounds. Here, we merely present color-coded contours of
$\langle\sigma v\rangle|_{v\to 0}$ in units of $10^{-24} cm^3/sec$.
The red-, yellow- and green-shaded regions will typically lead to observable
levels of gamma-ray or antimatter detection rates, if higgsino-like WIMPs dominate
the CDM relic density. However, in scenarios like the PQMSSM with 
mixed axion-WIMP dark matter, but with axion domination, these rates will be suppressed 
due to the low halo abundance of higgsino-like WIMPs.
In other PQMSSM cases where the axino $\ta$ is the LSP\cite{axinoLSP}, 
then the higgsino-like WIMPs would all
have decayed to relic axinos, and no direct  or indirect detection signals would be seen
(although detection of relic axions would still be possible).
\FIGURE[tbh]{
\includegraphics[width=13cm,clip]{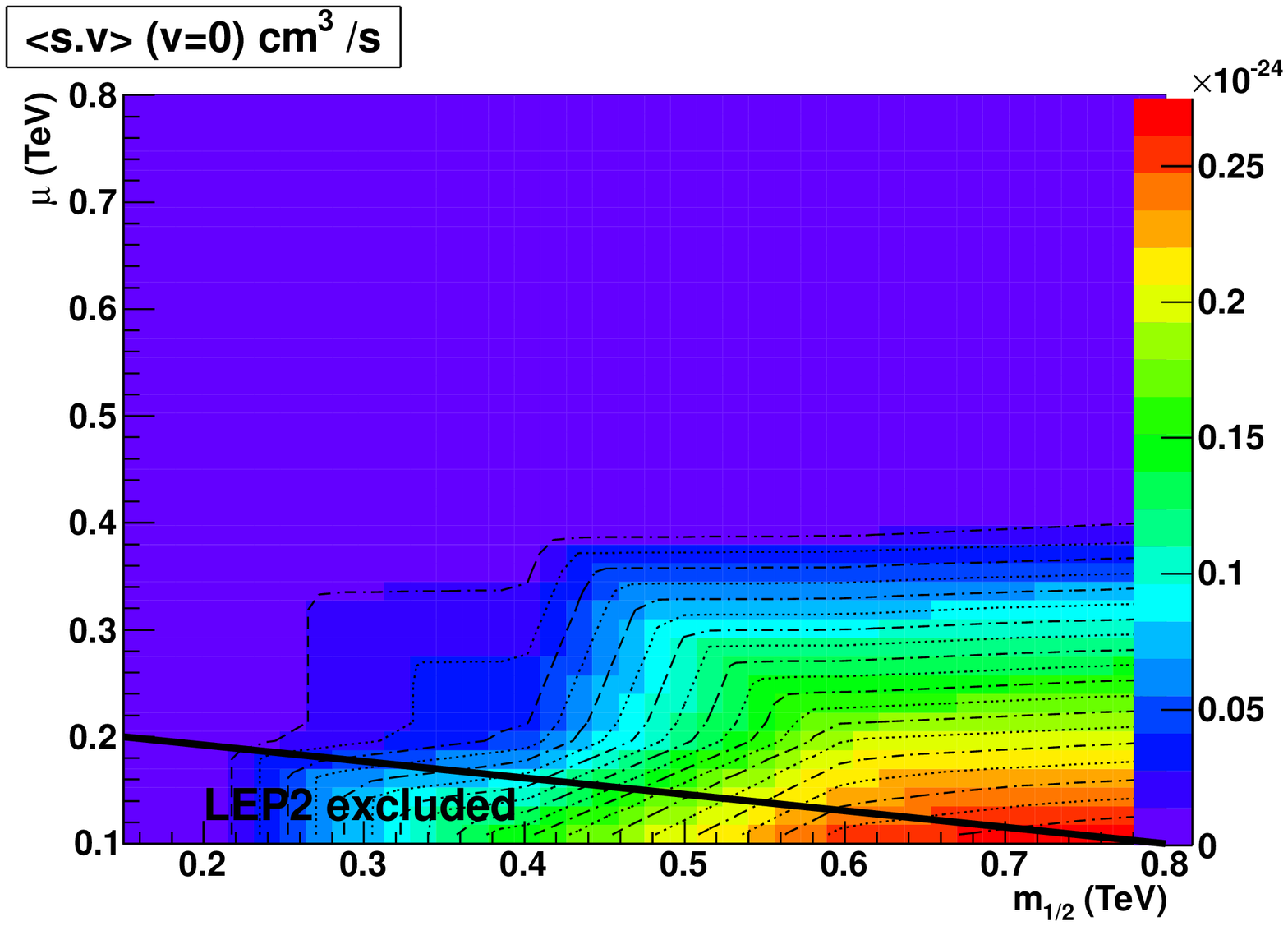}
\caption{Contours of $\langle\sigma v\rangle |_{v\to 0}$
in units of $10^{-24}$ $cm^3/sec$ in the $\mu\ vs.\ m_{1/2}$ plane with $m_A=800$ GeV, 
$m_0=5000$ GeV, $A_0=0$ and $\tan\beta =10$.
}
\label{fig:sdotv}}
%

\section{Higgsino-world scenario at the LHC}
\label{sec:lhc}

\subsection{Sparticle production at LHC7}

In the HW scenario, squarks and sleptons are assumed decoupled from collider physics.
In the limit of large scalar masses, the reach of LHC7\cite{lhc7} 
with 2 fb$^{-1}$ for gluino pair production is
to $m_{1/2}\sim 250$ GeV (corresponding to $m_{\tg}\sim 700$ GeV), while the
reach of LHC14 with 100 fb$^{-1}$ is to $m_{1/2}\sim 650$ GeV (corresponding 
to $m_{\tg}\sim 1400$ GeV)\cite{lhcreach}. Thus, for most of HW parameter space, 
gluino pair production will be below LHC sensitivity.
We then expect chargino/neutralino pair production to be the most promising
SUSY cross sections at LHC. 

In Fig. \ref{fig:lhc}, we show the dominant sparticle pair 
production cross sections in $fb$ for LHC7  
from the higgsino-world scenario versus $\mu$
for other model parameters as in Table \ref{tab:BM}. 
We adopt the computer code Prospino so that the results are valid at NLO
in QCD\cite{prospino}.
For low values
of $\mu$, we see that $\tw_1^\pm\tz_1$ and $\tw_1^\pm\tz_2$
are dominant, followed closely by $\tw_1^+\tw_1^-$ and $\tz_1\tz_2$
production. As $\mu$ moves beyond 300 GeV, the lighter
charginos and neutralinos become mixed gaugino-higgsino states, and
some of the cross sections drop rapidly.
Other potentially visible cross sections such as $\tz_2\tz_2$
are several orders of magnitude below these. 
The sum total of the reactions shown in Fig. \ref{fig:lhc}
agrees well with output for all SUSY reactions as generated by
Isajet as shown in Table \ref{tab:BM}, 
so these are indeed the dominant production reactions.
\FIGURE[tbh]{
\includegraphics[width=13cm,clip]{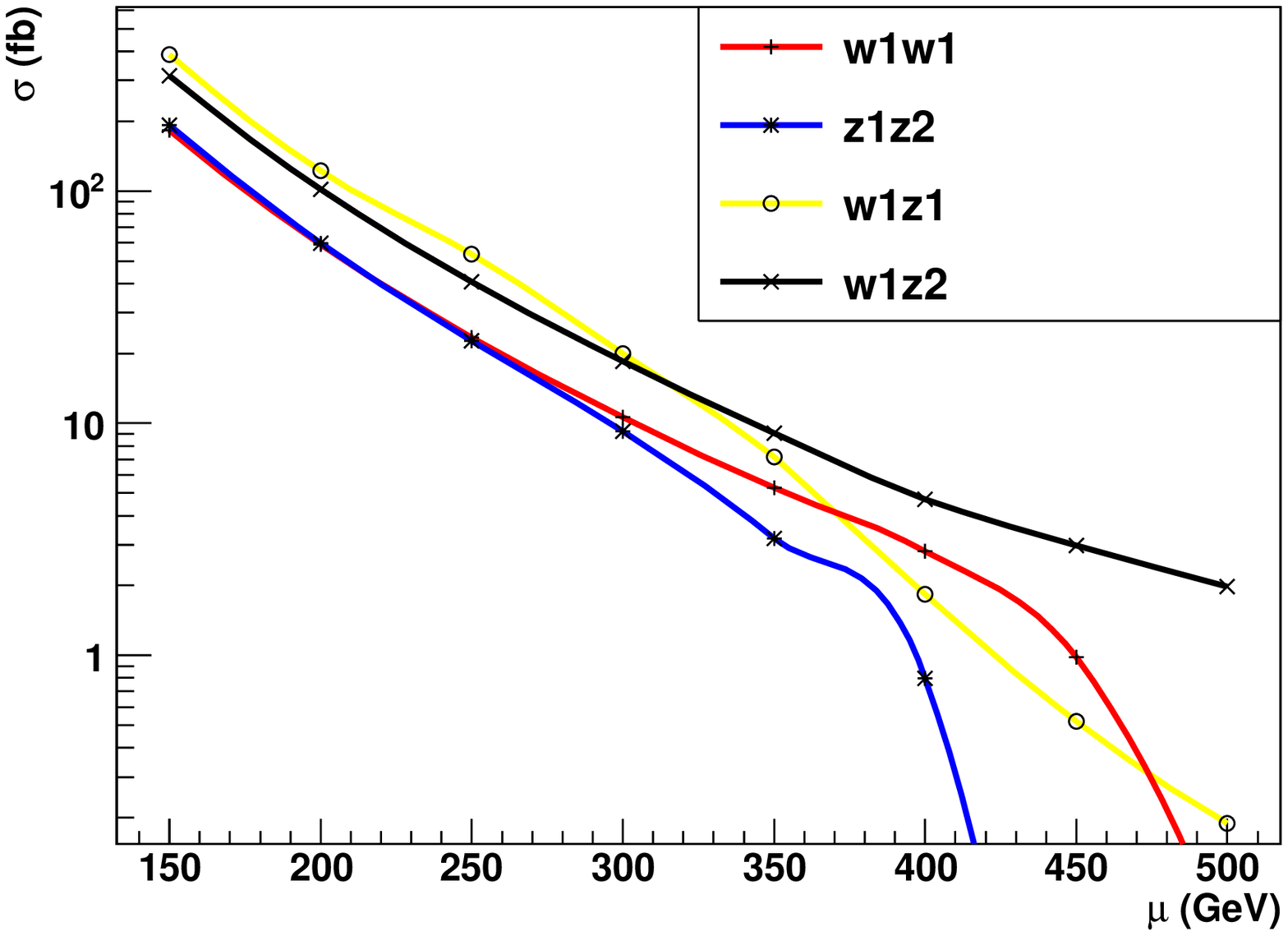}
\caption{Dominant chargino and neutralino production cross sections
versus $\mu$ at LHC7 for light higgsino-world 
SUSY scenario. Other parameters are fixed as in Fig. \ref{fig:mass}.
}
\label{fig:lhc}}

\subsection{Branching fractions and collider signatures}

The sparticle branching fractions can be read off from the Isajet
decay table for sparticle cascade decays\cite{cascade}.
For light charginos $\tw_1$, we find
\bi
\item $\tw_1^-\to\ell\bar{\nu}_\ell\tz_1$ at 11.1\% for each species 
$\ell =e$, $\mu$ or $\tau$,
\item $\tw_1^-\to d\bar{u}\tz_1$ at 33.3\% ,
\item $\tw_1^-\to s\bar{c}\tz_1$ at 33.3\% 
\ei
since the three-body chargino decays are dominated by the $W^*$ 
propagator. 

For $\tz_2$, we find typically
\bi
\item $\tz_2\to \ell^+\ell^-\tz_1$ at 3.5\% for each species 
$\ell =e$, $\mu$ or $\tau$,
\item $\tz_2\to \nu_\ell\bar{\nu}_\ell\tz_1$ at 21.5\% 
(summed over all neutrino species) and
\item $\tz_2\to q\bar{q}\tz_1$ at 68\% summed over all quark species.
\ei
In addition, the decay $\tz_2\to\tz_1\gamma$ occurs at an enhanced rate:
$0.8\%$ ($0.2\%$) for HW150 (HW300)\cite{z2z1g}.

By combining production cross sections with branching fractions,
we find that $\tw_1\tz_1$ production will lead to either
soft jets+MET (likely buried under QCD background(BG)) or a soft isolated lepton+MET
(likely buried under BG from direct $W$-boson production). 
Thus, we do not expect this reaction to lead to observable signatures.

The reaction $\tw_1^+\tw_1^-$ will lead to either 1) soft
jets+MET, 2) soft isolated lepton plus jets+MET  or
3) soft dilepton pair+MET. We expect each of these also to be buried beneath
SM backgrounds from QCD or vector boson pair production.

The reaction $\tz_1\tz_2$ production can lead to 1) soft jets+MET or 
2) soft, low invariant mass dilepton pairs+MET. The first of these is likely
buried beneath QCD background. The second of these has a chance at 
observability since the $m(\ell^+\ell^- )$ will be bounded by
$m_{\tz_2}-m_{\tz_1}$ and thus lead to a distinctive mass edge
upon a continuum background arising from $WW$ or $Z\gamma^*$ production, 
where $\gamma^*\to\ell^+\ell^-$. If the dilepton pair is at high $p_T$,
due to a highly boosted $\tz_2$, then we expect the dilepton pair to be highly
collimated in opening angle, and to appear rather distinctively compared to 
known backgrounds.

The reaction $\tw_1\tz_2$ production will lead to either 1) soft jets+MET, 
2) soft jets plus collimated soft dilepton+MET or 3) soft trileptons+MET.
The first of these is likely buried beneath QCD background. The second is possibly observable,
and should be present if the cleaner $\tz_1\tz_2\to \ell^+\ell^- +MET$ is found. The third 
case yields the venerable clean trilepton signature which has been evaluated for 
the Tevatron\cite{trilep} and LHC\cite{lhc3l,ullio}. While $W^*\gamma^*$ and $W^*Z^*\to 3\ell$
backgrounds proved most daunting for the Tevatron, at LHC the dominant background comes from $t\bar{t}$ production\cite{ullio}.

\subsection{Collimated dilepton +MET sigmature from $\tz_1\tz_2$ production}

We first investigate the $pp\to\tz_1\tz_2\to\ell^+\ell^-+MET$ 
signal against the following SM backgrounds:
\bi
\item $W^+W^-$ production (including $WW\to\tau^+\tau^-$),
\item $t\bar{t}$ production,
\item $\gamma^*\to\ell^+\ell^-$ (Drell-Yan) production,
\item $Z+jets$ with $Z\to\tau^+\tau^-$ (tau pair) production,
\item $\gamma^*Z$ production, where $Z\to\nu_\ell\bar{\nu}_\ell$ and
$\gamma^*\to \ell^+\ell^-$ or $\tau^+\tau^-$.
\ei
We generate sparticle production and decay events at parton level using Isajet\cite{isajet}
in Les Houches Event (LHE) format, and then feed the LHE files into 
Pythia\cite{pythia} for initial/final state radiation, hadronization
and underlying event. All backgrounds are generated with Pythia except $\gamma^*Z^*$ which
is generated by Madgraph/MadEvent\cite{madgraph}. The
collider events are then fed into the PGS toy detector simulation
program\cite{pgs}. 
Jets are found using an anti-$k_T$ jet finding algorithm with 
cone size $\Delta R=0.5$. Leptons are classified as isolated if they contain
less than 5 GeV hadronic activity in a cone of $\Delta R=0.2$ about the
lepton direction.

Since the leptons from $\tz_1\tz_2$ production are expected to be quite soft, 
we will focus initially upon the case of dimuon production, since muons
can be identified more easily than electrons at very low $p_T$.
Signal and background cross sections before and after cuts are listed 
in Table \ref{tab:SBG}.
We first require:
\bi
\item two opposite-sign muons: one with $p_T(\mu_1)>15$ GeV and 
$|\eta(\mu_1 )|<0.9$ (central region), while the other has
$p_T(\mu_2 )>5$ GeV with $|\eta (\mu_2)|<2.4$.
\ei
To reduce the large background from  Drell-Yan dimuon production, 
we next impose
\bi
\item $MET>25$ GeV,
\ei
since $MET$ in the DY case only arises from particles lost along the beam-line or cracks, 
or from energy mis-measurement, mainly from hadron radiation.
There is also a large background from $t\bar{t}$ production of dimuons and a background from single top in association with a W-boson which has a hard b-jet, but 
this always comes along with two hard $b$-jets from the $t\to bW$ decays. Thus, we also
require the number of jets
\bi
\item $n(jets)=0$,
\ei
where jets are identified as a cluster of hadrons with $p_T(jet)>15$ GeV, $|\eta (jet)|<2.4$
At this stage, the largest background comes from $W^+W^-$ production, which yields
a continuum distribution in dimuon invariant mass $m(\mu^+\mu^- )$, 
whilst the signal dimuons are restricted to $m(\mu^+\mu^- )<m_{\tz_2}-m_{\tz_1}$
which is just 16.2 GeV for HW150. Thus, we require
\bi
\item $m(\mu^+\mu^- )<20$ GeV .
\ei
Signal and BG after these cuts are listed for HW150 in Table \ref{tab:SBG}. 

From Table \ref{tab:SBG}, we see that the dimuon signal comes about 30\% from $\tz_1\tz_2$ 
production, and about 70\% from $\tw_1\tz_2$ production. In the latter case, 
the $\tw_1$ usually decays to $q\bar{q}'\tz_1$ but with very low energy release, 
which sometimes escapes the ``no-jet'' cut. 
Meanwhile, the dominant remaining background comes from 
tau-pair, Drell-Yan and $W^+W^-$ production.
The remaining signal is 0.43 fb, while the summed SM background is $\sim 11.9$ fb.
The $5\sigma$ discovery cross section for LHC7, assuming 10 fb$^{-1}$ of
integrated luminosity is 5.45 fb, so that the case of HW150 is far below this limit.
%
\begin{table}\centering
\begin{tabular}{lcc}
\hline
process & $\sigma$ (fb) & $\sigma$ (after cuts, fb) \\
\hline
$\tw_1\tz_2$   & 313 & 0.3 \\
$\tz_1\tz_2$   & 192 & 0.13 \\
$\gamma^*\to \mu^+\mu^-$ (DY)   & $1.1\times10^{6}$ & 4\\
$W^+W^-\to\mu^+\mu^-$       & $235.5$ & 2.3 \\
$\gamma^*Z\to\mu^+\mu^-\nu_i\bar{\nu}_i$    & 6.8 & 0.3 \\
$\gamma^*,Z\to \tau^+\tau^-\to\mu^+\mu^-$ & $1.5\times10^{4}$ & 5  \\
$t\bar{t}\to\mu^+\mu^-$     & $8.9\times10^{4}$ & $<0.3$ \\
\hline
\hline
\end{tabular}
\caption{Signal and BG cross sections in fb 
before and after cuts at LHC7. The signal rates are for higgsino-world benchmark point HW150. 
Each background process requires $p_T(\mu )>5 $ GeV.
}
\label{tab:SBG}
\end{table}

The inclusive muon $p_T$ distribution before cuts is shown in Fig.~\ref{fig:pt}
for the HW150 benchmark.
Here, we see that the spectrum from HW150 benchmark is very soft, with the bulk
of the distribution below 15 GeV. Thus, few of the signal events escape even the first
cut listed above on $p_T(\mu )>15$ GeV. 
The signal rates for HW150 after cuts only corresponds to four events in 10 fb$^{-1}$
of integrated luminosity, while SM background lies at the $\sim 120$ event level.
\FIGURE[tbh]{
\includegraphics[width=13cm,clip]{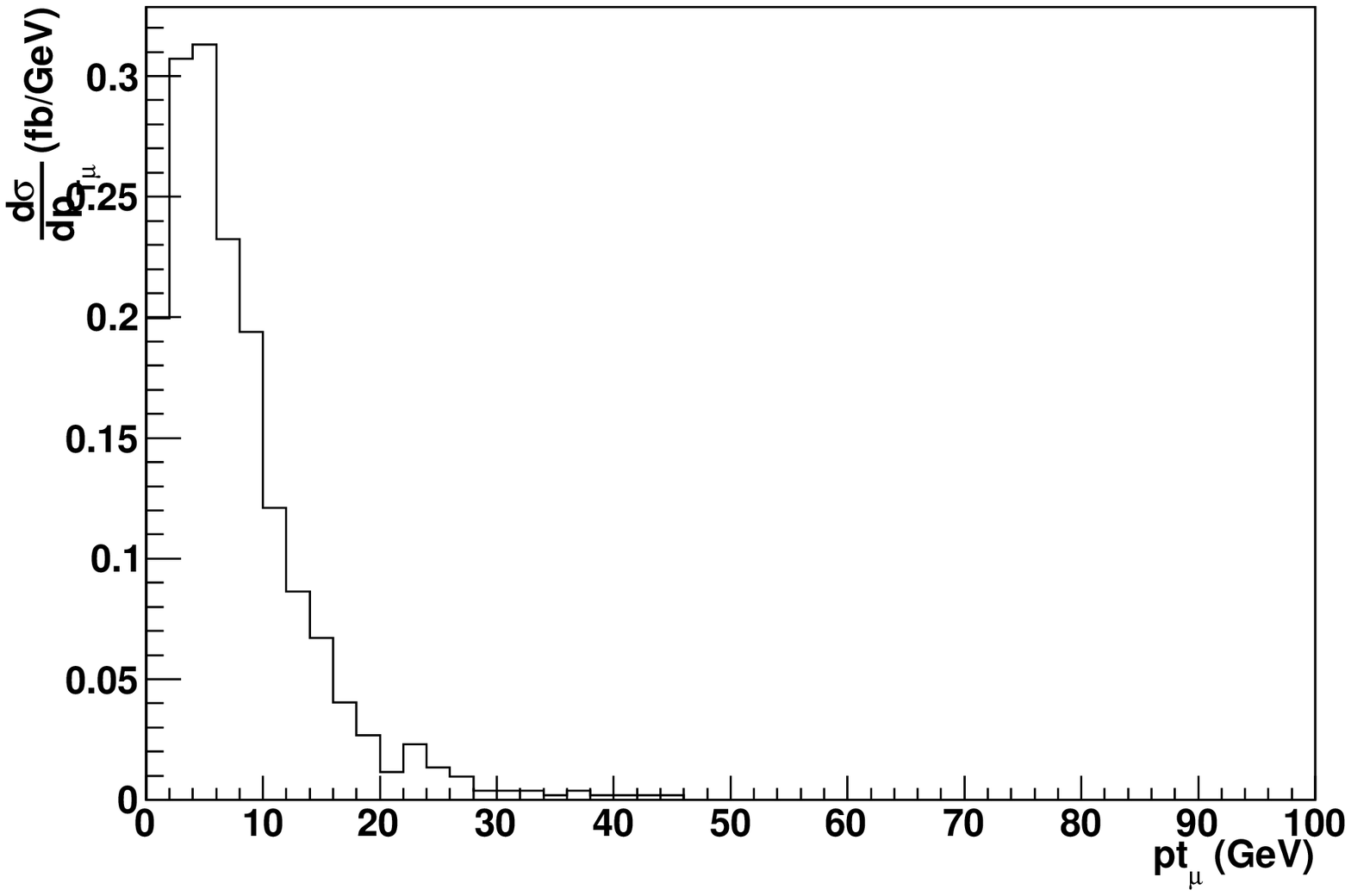}
\caption{Distribution in $p_T(\mu )$ from $pp\to\tz_1\tz_2\to\mu^+\mu^- +\eslt$
events at LHC from higgsino-world benchmark point HW150.
}
\label{fig:pt}}

Likewise, trilepton signatures from $\tw_1^\pm\tz_2$ production yield a very soft
isolated lepton spectrum, and are also difficult to extract at an observable level.
The search for jets$+MET$ from higgsino pair production also yields a
very soft jet and MET spectrum, and is difficult to extract from prodigious SM backgrounds.

Thus, higgsino-world SUSY seems capable of eluding standard SUSY searches via isolated 
multi-leptons. The key feature of the HW scenario is that possibly observable levels of
dimuon and trimuon production can occur, but at very low $p_T$ levels. Our studies then
motivate our experimental colleagues to push for dimuon and trimuon analyses at the very lowest $p_T(\mu )$ levels in order to extract a possible signal.

\section{Prospects for ILC or a muon collider}
\label{sec:ilc}

We have seen that the LHC has essentially no reach for the HW SUSY scenario
due to a very soft spectrum of observable sparticle decay products.
However, we have seen that the HW scenario mainly occurs for $\mu \alt 250$ GeV 
(for $m_{1/2}\alt 1$ TeV), which also corresponds to $m_{\tw_1}\alt 250$ GeV. 
Contours of $2m_{\tw_1}$ are shown in Fig. \ref{fig:w1}, where we see that 
the region with $2m_{\tw_1}\alt 500$ GeV covers almost all of
HW parameter space (compare against Fig. \ref{fig:vH}).
For this mass range, 
chargino pair production, and also $\tz_1\tz_2$ production, should be within
range of the proposed international Linear Collider (ILC), which is proposed
to operate initially at an energy $\sqrt{s}=500$ GeV. Chargino
pair production would also be accessible to higher energy $e^+e^-$ colliders
like CLIC, or a muon collider (MC) operating in the TeV regime.
\FIGURE[tbh]{
\includegraphics[width=13cm,clip]{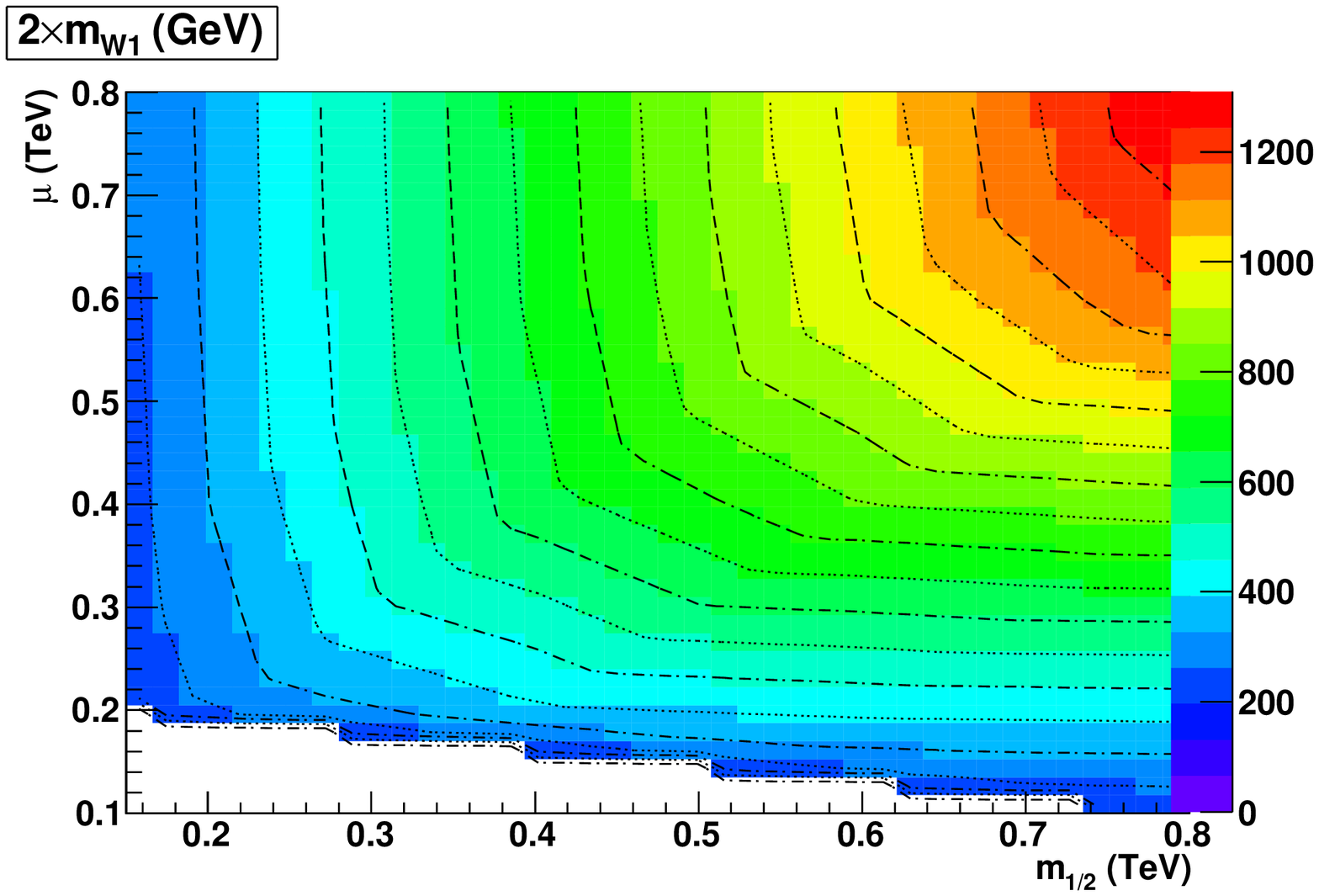}
\caption{Contours of $2m_{\tw_1}$ in the $\mu$ versus $m_{1/2}$ 
parameter plane for other SUSY parameters as in HW150 scenario.
}
\label{fig:w1}}

In Fig. \ref{fig:ilc}, we show the cross sections for $e^+e^-\to\tw_1^+\tw_1^-$ 
and $e^+e^-\to\tz_1\tz_2$ using SUSY parameters as in the HW150 benchmark, but
with $\mu$ varying from 100-250 GeV. The variation in $\mu$ causes $m_{\tw_1}$ to vary,
and in fact $m_{\tw_1}\sim \mu$, so that our results are plotted versus the 
more physical $m_{\tw_1}$ value. We take $\sqrt{s}=500$ GeV.
We see that over most of HW parameter space, the chargino pair production cross section is
in the several hundred fb range, until $m_{\tw_1}$ approaches the kinematic limit
for pair production. Chargino pair production will be signaled at ILC or MC
by 1) soft multijet $+\esl$ production, 2) soft isolated lepton plus jets $+\esl$
production and 3) dilepton $+\esl$ production, depending on whether the charginos decay
leptonically or hadronically. These signatures should be easily visible against 
SM backgrounds such as $WW$ production via distributions such as ``missing mass'':
$\msl =\sqrt{\esl^2-\psl^2}$\cite{bmt}. In addition, SM backgrounds such as dilepton
or dijet production from the $\gamma\gamma$ initial state will contain 
energy depositions all in the same plane, while the SUSY signal will contain
acoplanar events.
Thus, the HW scenario should be easily visible at ILC, or a higher energy muon collider, 
even though it is difficult to see at LHC.
\FIGURE[tbh]{
\includegraphics[width=13cm,clip]{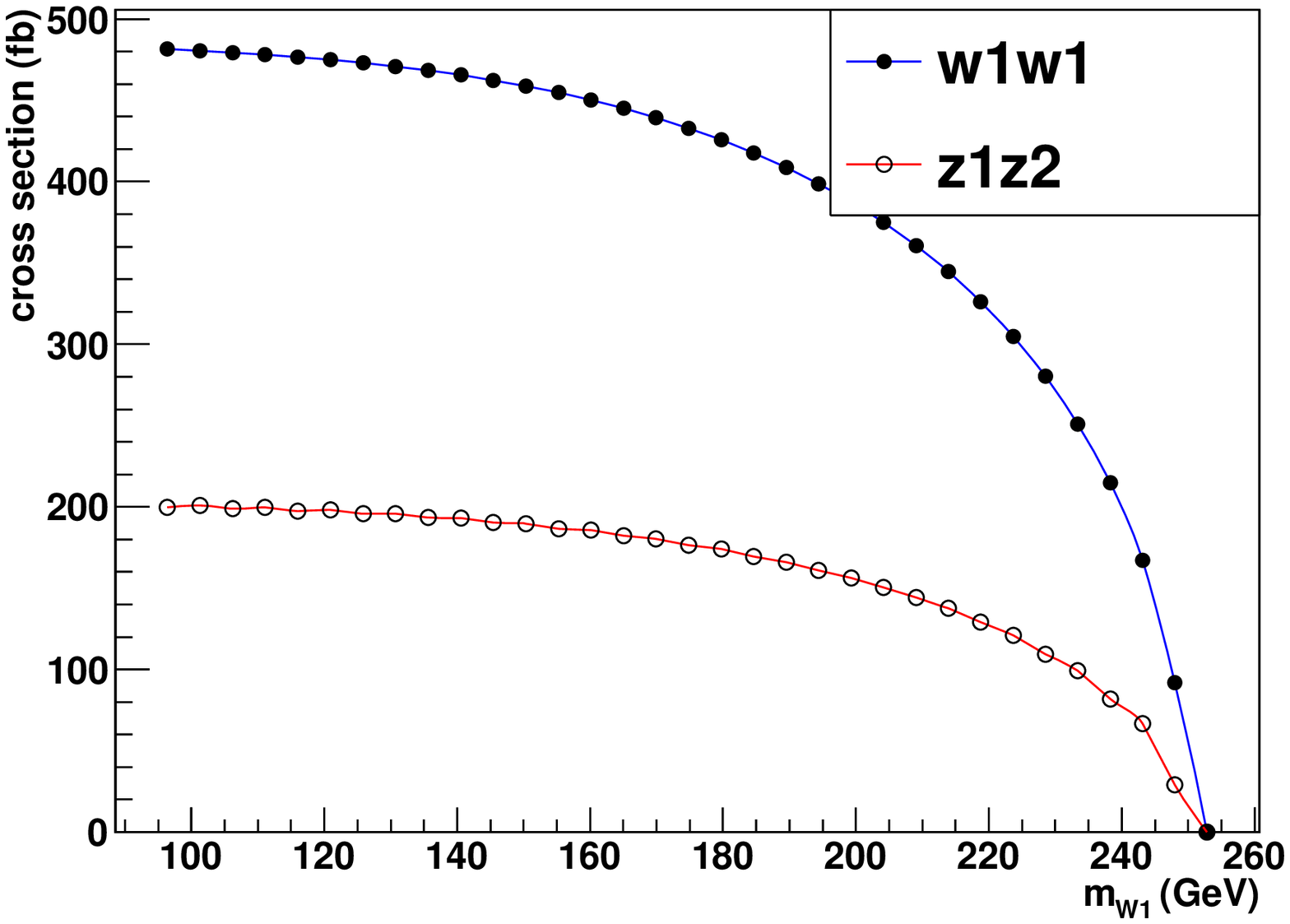}
\caption{Cross sections for chargino pair production and neutralino
pair production versus $m_{\tw_1}$ at a $\sqrt{s}=500$ GeV ILC or MC collider.
We take SUSY parameters as in Fig. \ref{fig:lhc}, and vary $\mu$
to give variation in $m_{\tw_1}$. 
}
\label{fig:ilc}}

A distinctive feature of the HW scenario is that the $\tw_1$, $\tz_1$ and $\tz_2$
are all mainly higgsino-like, whereas in models such as mSUGRA, these states
are almost always gaugino-like. In Ref. \cite{bmt}, it is shown that for
wino-like $\tw_1$ and $\tz_2$, the $\tw_1^+\tw_1^-$ and $\tz_1\tz_2$
production cross sections are steeply increasing functions of the
electron beam polarization $P_L(e^-)$ (where $P_L(e^-)\sim -1$ corresponds
to pure right-polarized $e^-$, $P_L(e^-)=+1$ corresponds to pure left-polarized 
$e^-$, and $P_L(e^-)=0$ corresponds to unpolarized $e^-$ beams).
In Fig. \ref{fig:epol}, we plot the $e^+e^-\to\tw_1^+\tw_1^-$ and $\tz_1\tz_2$
cross sections versus $P_L(e^-)$ for the HW150 benchmark. In the HW scenario,
$\tw_1^+\tw_1^-$ production only increases by a factor of $\sim 3.5$ as 
$P_L(e^-)$ varies from -1 to +1, whereas in mSUGRA it typically
increases by factors of about 100\cite{bmt}. In addition, the
$\tz_1\tz_2$ cross section for HW150 is nearly flat versus $P_L(e^-)$, 
while in mSUGRA, it is typically increasing by factors of 20-30.
Thus, variability of the SUSY production cross sections
versus beam polarization will quickly allow one to extract
much of the gaugino/higgsino content of the charginos/neutralinos
which are accessible to an ILC with adjustable beam polarization. 
\FIGURE[tbh]{
\includegraphics[width=13cm,clip]{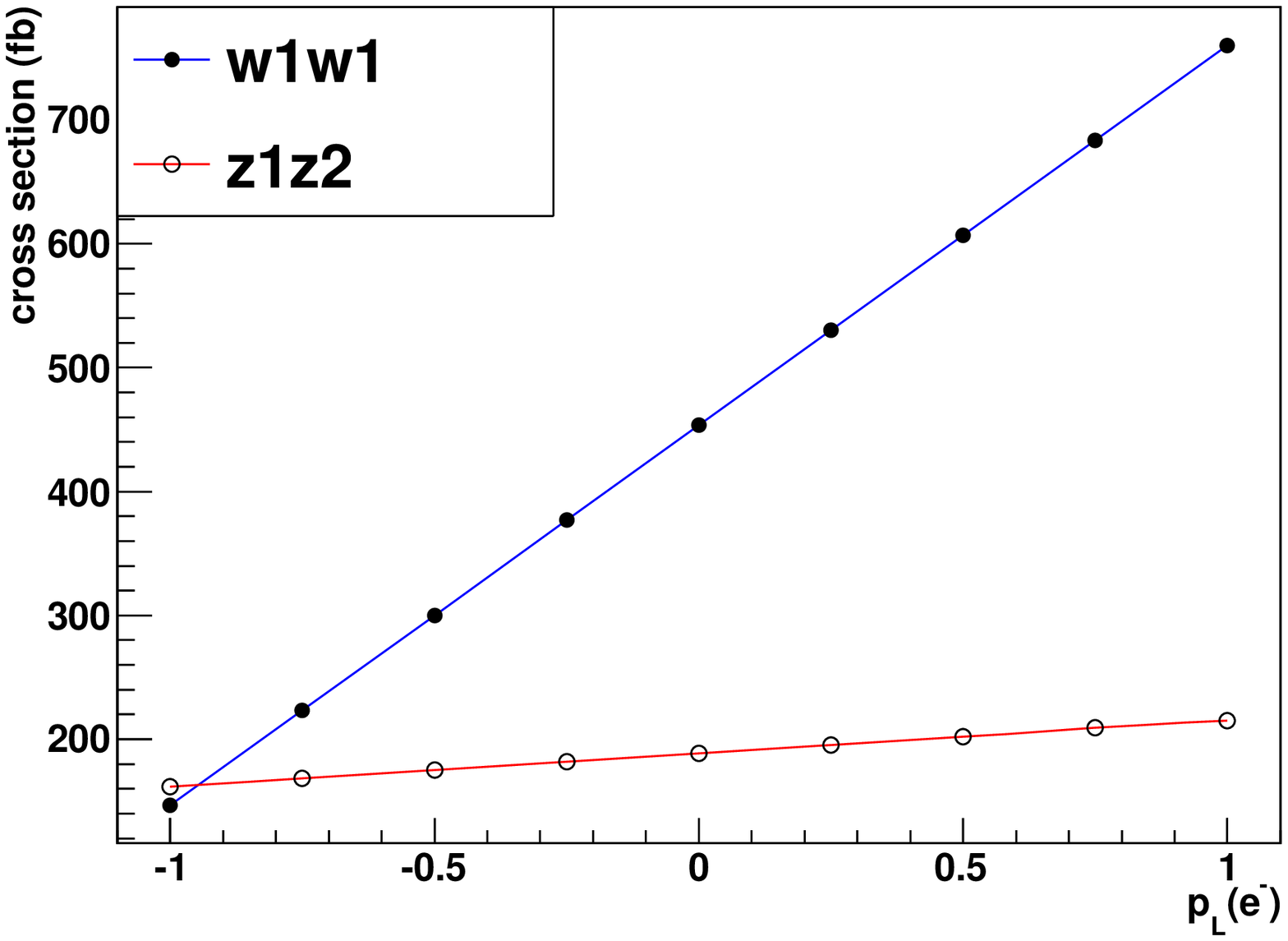}
\caption{Cross sections for chargino pair production and neutralino
pair production versus $P_L(e^- )$ at a $\sqrt{s}=500$ GeV ILC collider.
We take SUSY parameters as in HW1, with $\mu =150$ GeV.
}
\label{fig:epol}}
%

\section{Summary and conclusions}
\label{sec:conclude}

The higgsino-world SUSY scenario with multi-TeV scalars, 
$\mu\alt 250$ GeV and intermediate scale gauginos is
 very appealing in that it can reconcile a decoupling
solution to the SUSY flavor, CP, $p$-decay and gravitino problems
with apparently low levels of naturalness or electroweak
fine-tuning. 
The scenario is characterized by a mass hierarchy
$|\mu |\ll m_{1/2}\ll m_0$, where $m_0$ is the GUT scale 
mass of matter scalars. The HW scenario is most easily realized in models
with non-universal Higgs masses, where the weak scale values of $\mu$ and
$m_A$ are taken as free parameters.
In the HW scenario, the $\tw_1$, $\tz_1$ and $\tz_2$
states are all light with mass $\alt 250$ GeV, and dominantly higgsino-like.
The remaining sparticles may well be heavy and inaccessible to LHC searches.

The standard thermal abundance of higgsino-like $\tz_1$ particles
is well below WMAP-measured values. However, in appealing cosmological
scenarios such as those containing TeV-scale scalar fields such as moduli, 
or in scenarios with mixed axion-$\tz_1$ cold dark matter, the
neutralino abundance can be easily pushed up into the measured range.
If this is so, then there are excellent prospects for direct or indirect 
detection of higgsino-like relic WIMPs, and we expect experiments such
as Xenon-100 or Xenon-1-ton to fully explore this possibility.
Alternatively, in DM models such as the mixed $a\tz_1$ scenario\cite{blrs}, it is 
also possible to tune PQ parameters such that the WIMP abundance remains tiny, while
the bulk of CDM is comprised of axions. Thus, the HW scenario will not be
completely excludable by direct or indirect WIMP search experiments if no signals
for WIMPs are seen.

At the LHC, gluino and squark production may be suppressed
by large values of $m_{\tg}$ and especially $m_{\tq}$. 
The $\tw_1\tz_1$, $\tw_1\tz_2$, $\tz_1\tz_2$ and $\tw_1^+\tw_1^-$ 
production reactions are then dominant,
but are difficult to detect at LHC due to the small $\tw_1 -\tz_1$ and
$\tz_2-\tz_1$ mass gaps, which lead to very soft visible particle production.
The reaction $pp\to\tz_1\tz_2$ may lead to tightly collimated OS/SF
dilepton pairs, although calculations of signal and background after simple cuts
indicate these occur at unobservable levels.
Trileptons from $\tw_1\tz_2$ production 
are also difficult to see due to the soft spectrum of isolated leptons.
Our studies should motivate our experimental colleagues to push for di- and tri-muon
analyses at the very lowest levels of $p_T(\mu )$ which are possible.

A linear $e^+e^-$ collider such as ILC  or a $\mu^+\mu^-$ collider operating 
with $\sqrt{s}\sim 0.5-1$ TeV
should be able to make a thorough search for the HW scenario. 
If HW SUSY is discovered at ILC, then it should be possible
to extract the gaugino/higgsino content of the $\tw_1$, $\tz_2$ and $\tz_1$ states using
various kinematic and angular distributions along with beam polarization.
Thus, the HW scenario provides a concrete realization of a SUSY construct
which may well remain hidden from LHC and dark matter searches, 
but which is fully testable at a TeV-scale lepton collider.

\acknowledgments

This work was supported in part by the U.S. Department of Energy under grant No.~DE-FG02-04ER41305 and DE-FG02-95ER40896. 
VB thanks the KITP-Santa Barbara for its hospitality.


%

\end{document}